\documentclass[superscriptaddress,twocolumn]{revtex4}
\usepackage{amssymb}
\usepackage[tbtags]{amsmath}
\usepackage{graphicx}
\usepackage{epsfig,graphicx,times}

\begin{document}

\title{Macroscopic Einstein-Podolsky-Rosen pairs in superconducting circuits}

\author{L.F. Wei}
\affiliation{Frontier Research System, The Institute of Physical
and Chemical Research (RIKEN), Wako-shi, Saitama, 351-0198, Japan}
\affiliation{Institute of Quantum Optics and Quantum Information,
Department of Physics, Shanghai Jiaotong University, Shanghai
200030, P.R. China}

\author{Yu-xi Liu}
\affiliation{Frontier Research System, The Institute of Physical
and Chemical Research (RIKEN), Wako-shi, Saitama, 351-0198, Japan}

\author{Markus J. Storcz}
\affiliation{Frontier Research System, The Institute of Physical
and Chemical Research (RIKEN), Wako-shi, Saitama, 351-0198, Japan}
\affiliation{Physics Department, ASC, and CeNS,
Ludwig-Maximilians-Universit\"at, Theresienstrasse 37, 80333
M\"unchen, Germany}

\author{Franco Nori}
\affiliation{Frontier Research System, The Institute of Physical
and Chemical Research (RIKEN), Wako-shi, Saitama, 351-0198, Japan}
\affiliation{Center for Theoretical Physics, Physics Department,
CSCS, The University of Michigan, Ann Arbor, Michigan 48109-1040,
USA}

\date{\today }

\begin{abstract}

We propose an efficient approach to prepare
Einstein-Podolsky-Rosen (EPR) pairs in currently existing
Josephson nanocircuits with capacitive couplings. In these fixed
coupling circuits, two-qubit logic gates could be easily
implemented while, strictly speaking, single-qubit gates cannot be
easily realized. For a known two-qubit state, conditional
single-qubit operation could still be designed to evolve only the
selected qubit and keep the other qubit unchanged; the rotation of
the selected qubit depends on the state of the other one.
These conditional single-qubit operations allow to
deterministically generate the well-known Einstein-Podolsky-Rosen
pairs, represented by EPR-Bell (or Bell) states. Quantum-state
tomography is further proposed to experimentally confirm the
generation of these states. The decays of the prepared EPR pairs
are analyzed using numerical simulations. Possible application of
the generated EPR pairs to test Bell's Inequality is also
discussed.

\vspace{0.3cm} PACS number(s): 03.67.Mn, 03.65.Wj, 85.25.Dq.

\end{abstract}

\maketitle

\section{Introduction}

Quantum mechanics (QM) is a very successful theory. It has solved
many physical mysteries in both macroscopic superconductivity and
microscopic particles. Still, laboratory studies of its conceptual
foundation and interpretation continue to attract much attention.
One of the most important examples is the well-known
Einstein-Podolsky-Rosen (EPR) ``paradox", concerning the
completeness of QM.
Based on a {\it gedanken} experiment, Einstein, Podolsky and Rosen
(EPR) claimed~\cite{EPR35} that QM is incomplete and that
so-called ``hidden variables" should exist. This is because a
two-particle quantum system might be prepared in a correlated
(i.e., entangled) state, even though the two particles are
spatially separated by a large distance and without any direct
interaction. A measurement performed on one of the particles
immediately changes the state (and thus the possible physical
outcome) of the other particle. This ``paradox" leads to much
subsequent, and still on-going, researches.
Bell proposed~\cite{Bell64} an experimentally testable inequality
to examine the existence of the hidden variables: if this
inequality is violated, then there are no so-called local ``hidden
variables", and thus quantum mechanical predication of existing
quantum non-local correlations (i.e., entanglement) is sustained.

During the past decades, a number of interesting
experiments~\cite{asp82} using entangled photon pairs have been
proposed and carried out to investigate the two-particle non-local
correlations. These experiments showed that Bell's inequality (BI)
could be strongly violated, and agreed with quantum mechanical
predictions. Yet, one of the essential loopholes in these optical
experiments is that the required EPR pairs were probabilistically
generated in a small subset of all photons created in certain
spontaneous processes.
Thus, it is necessary to study two-particle entanglement in
different, e.g., massive or macroscopic systems, instead of
fast-escaping photons. Expectably, the EPR pairs between these
massive ``particles" can be deterministically prepared.
Theoretical proposals include those with e.g., neutral
Kaons~\cite{Barmon99}, Rydberg atoms~\cite{Oliver87}, ballistic
electrons in semiconductors~\cite{Ionicioiu01}, and trapped
ions~\cite{wei04-epl-1}.
Experimentally, two Rydberg atoms had been first entangled to form
EPR pair in a high $Q$ cavity by the exchange of a single
photon~\cite{Raimond97}. Later, by exchanging the quanta of the
common vibrational mode, EPR correlations with ultralong lifetime
(e.g., up to $5$ microsecond) had been generated between a pair of
trapped cold ions~\cite{Blatt04}. Consequently, violations of BI
have been experimentally verified with the EPR correlations
between either the two ions~\cite{Rowe01}, or an atom and a
photon~\cite{Monroe04}.

Recent developments of quantum manipulation in coupled Josephson
systems~\cite{Pashkin03,martinis05} allow to experimentally
investigate the quantum correlations between two macroscopic
degrees of freedom in a superconducting nano-electronic
device~\cite{Berman05}.
Proposals have been made for producing quantum entanglement
between two superconducting qubits, e.g., indirectly coupled by
sequentially interacting with a current-biased information
bus~\cite{Blais03,wei04-epl-2}, coupled
inductively~\cite{Tanaka01,makhlin99}, and coupled via either a
cavity mode~\cite{He03} or a large Josephson
junction~\cite{You03}. By introducing an effective dynamical
decoupled approach, we have shown~\cite{wei04-arXiv} that the BI
could also be tested with superconducting qubits, even if the
interaction between them is fixed. The robustness of the scheme
proposed in Ref.~\cite{wei04-arXiv} is better suited for weak
interbit couplings, e.g., when the ratio of the interbit-coupling
energy $E_{m}$ and the Josephson energy $E_J$ of the qubit is
small. In this paper, for {\it an arbitrary interbit coupling}
strength, we discuss how to prepare the EPR correlations, i.e.,
deterministically generate and tomographically measure the
well-known EPR-Bell (or Bell) states:
\begin{equation}
|\psi _{\pm }\rangle =\frac{1}{\sqrt{2}}(|00\rangle \pm
|11\rangle),\,\,|\phi _{\pm }\rangle
=\frac{1}{\sqrt{2}}(|01\rangle \pm |10\rangle),
\end{equation}
in a capacitively coupled Josephson circuit. Its possible
application to directly test the EPR paradox is also discussed.

The outline of the paper is as follows. In Sec. II, a few
elementary quantum operations are proposed to deterministically
manipulate two charge qubits coupled capacitively. Some of them
only evolve a selected qubit and leave the remaining one
unaffected. These operations are not strictly single-qubit gates
(just {\it conditional} single-qubit operations), as the rotation
of the selected qubit depends on the state of the other qubit. By
making use of these operations, in Sec. III, we propose a two-step
approach to deterministically generate the EPR pairs from the
circuit's ground state $|\psi (0)\rangle =|00\rangle$. Further, we
discuss how to experimentally confirm the generation of EPR pairs
by tomographic measurements.
In Sec. IV, considering the existence of typical voltage-noises
and $1/f$-noise, we numerically analyze the decays of the prepared
EPR correlations within the Bloch-Redfield formalism~\cite{AK64}.
In Sec. V, we discuss the possibility of testing BI with the
generated EPR pairs. Conclusions and discussions are given in Sec.
VI.

\section{Manipulations of two capacitively coupled Josepshon charge qubits}

We consider the two-qubit nano-circuit sketched in Fig. 1, which
is similar to that in recent
experiment~\cite{Pashkin03,Yamamoto03}.
\begin{figure}[tbp]
\vspace{2.2cm}
\includegraphics[width=13.6cm]{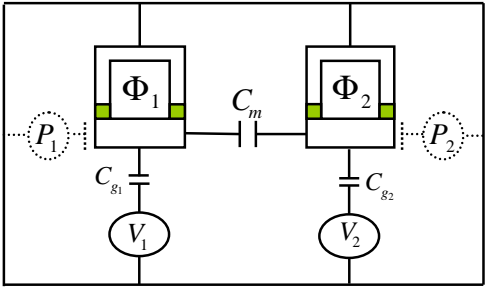}
\vspace{-8cm} \caption{(Color online) Two capacitively-coupled
SQUID-based charge qubits. The quantum states of two Cooper-pair
boxes (i.e., qubits) are manipulated by controlling the applied
gate voltages $V_1,\,V_2$ and external magnetic fluxes
$\Phi_{1},\,\Phi_{2}$ (threading the SQUID loops). $P_{1}$ and
$P_{2}$ (dashed line parts) read out the final qubit states.}
\end{figure}
Two superconducting quantum interference device (SQUID) loops with
controllable Josephson energies produce two Cooper-pair boxes,
fabricated a small distance apart~\cite{Pashkin03,Yamamoto03}) and
coupled via the capacitance $C_{m}$. The Hamiltonian of the
circuit reads
\begin{equation}
\hat{H}=\sum_{j=1,2}[E_{C_j}(\hat{n}_j-n_{g_j})^2-E_{J}^{(j)}\cos\hat{\theta}_j]
+E_m\prod_{j=1}^2(\hat{n}_j-n_{g_j}),
\end{equation}
in the charge basis. Here, the excess Cooper-pair number operator
$\hat{n}_j$ and phase operator $\hat{\theta}_j$ in the $j$th box
are conjugate: $[\hat{\theta}_j,\hat{n}_k]=i\delta_{jk}$.
$E_{C_{j}}=4e^{2}C_{\Sigma _{k}}/C_{\Sigma },\,j\neq k=1,2$ and
$E_{J}^{(j)}=2\varepsilon _{J_j}\cos( \pi \Phi _{j}/\Phi _{0})$
are the charging and Josephson energies of the $j$th box.
$E_{m}=4e^{2}C_{m}/C_{\Sigma }$ is the coupling energy between the
boxes. Above, $\varepsilon _{J_j}$ and $C_{\Sigma _{j}}$ are the
Josephson energy of the single-junction and the sum of all
capacitances connected to the $j$th box, respectively. Also,
$C_{\Sigma }=C_{\Sigma _{1}}C_{\Sigma _{2}}-C_{m}^{2}$ and
$n_{g_j}=C_{g_{j}}V_{j}/(2e)$. $e$ is the electron charge and
$\Phi _{0} $ the flux quantum.
The circuit works in the charge regime with $k_{B}T\ll
\varepsilon_{J_j}\ll E_{C_j}\ll \Delta $\thinspace, wherein
quasi-particle tunnelling and excitation are effectively
suppressed and the number $n_{j}$\thinspace (with
$n_{j}=0,1,2,...$) of Cooper-pairs in the $j$th boxe is a good
quantum number. Here, $k_{B},\,T,\,\Delta$, and
$2\varepsilon_{J_j}$ are the Boltzmann constant, temperature,
superconducting gap, and maximal Josephson energies of the $j$th
Cooper-pair box, respectively.

Following Refs.~\cite{Pashkin03,Yamamoto03}, the dynamics of the
system near the co-resonance point (where $n_{g_1}=n_{g_2}=1/2$)
can be effectively restricted to the subspace $\Xi$ spanned by
only the four lowest charge states: $|00\rangle, |10\rangle,
|01\rangle$ and $|11\rangle $, and thus the above Hamiltonian can
be simplified to
\begin{equation}
\hat{H}=\sum_{j=1,2}\frac{1}{2}\left[ E_{C}^{(j)}\sigma
_{z}^{(j)}-E_{J}^{(j)}\sigma _{x}^{(j)}\right]+E_{12}\,\sigma
_{z}^{(1)}\sigma _{z}^{(2)},
\end{equation}
with $E_{12}=E_m/4$, and
$E_{C}^{(j)}=E_{C_{j}}(n_{g_j}-1/2)+E_{m}( n_{g_k}/2-1/4),\,j\neq
k=1,2$. The pesudospin operators are defined as $\sigma
_{z}^{(j)}=|0_j\rangle \langle 0_j|-|1_j\rangle \langle 1_j|$ and
$\sigma _{x}^{(j)}=|0_j\rangle \langle 1_j|+|1_j\rangle \langle
0_j|$. Here, the subindex $j$ (or $k$) is introduced to label the
state of the $j$th (or $k$th) qubit. For example, $|0_j\rangle$
refers to the logic state of the $j$th qubit is ``0". For
simplicity, the subindexes in a two-qubit state $|mn\rangle$ (with
$m,n=0,1$) are omitted, and $m$ ($n$) usually (except when
indicated otherwise) refers to the state
$|m\rangle$\,($|n\rangle$) of the first (second) qubit.

Obviously, the interbit-coupling energy $E_{12}=E_m/4$ is
determined by the coupling capacitance $C_m$ and therefore is
fixed by fabrication, i.e., not controllable. However, $E_C^{(j)}$
and $E_J^{(j)}$ can be controlled by adjusting the applied
gate-voltages $V_j$ and fluxes $\Phi_j$, respectively. Although
any evolution of this two-qubit system is solvable and can be
expressed by a $4\times 4$ matrix in the subspace $\Xi$, we prefer
certain relatively simple quantum operations by properly setting
the above controllable parameters to conveniently engineer
arbitrary quantum states. These operations are summarized in the
following three subsections.

\subsection{Operational delay}

First, we assume the circuit stays in the parameter settings such
that $E_{C}^{(j)}=E_J^{(j)}=0$, until any operation is applied to
it. Thus, during the operational delay $\tau$, the circuit evolves
under the Hamiltonian $\hat{H}_{\rm int}=E_{12}\,\sigma
_{z}^{(1)}\sigma _{z}^{(2)}$, i.e., undergoes a free
time-evolution
\begin{eqnarray}
\hat{U}_0=\left(
\begin{array}{cccc}
e^{-i\alpha_0}&0&0&0\\
0&e^{i\alpha_0}&0&0\\
0&0&e^{i\alpha_0}&0\\
0&0&0&e^{-i\alpha_0}
\end{array}
\right),\alpha_0=\frac{E_{12}}{\hbar}\tau.
\end{eqnarray}
In this case, the Bell states in Eq. (1) will not evolve, once
they have been generated.

\subsection{Simultaneously evolving two qubits}

Due to the constant coupling, simultaneous operations on two
qubits are relatively easy. For example, if $n_{g_1}=n_{g_2}=1/2$
(i.e., at co-resonance point) and $E_J^{(1)}=E_J^{(2)}=E_J$, then
the circuit has the Hamiltonian $\hat{H}_{\rm
co}=-E_J(\sigma_x^{(1)}+\sigma_x^{(2)})/2+E_{12}\,\sigma_z^{(1)}\sigma_z^{(2)}$,
which produces the following time-evolution operator
\begin{eqnarray}
\bar{U}_{\rm co}=\frac{1}{2}\left(
\begin{array}{cccc}
a&b&b&c\\
b&a^*&c^*&b\\
b&c^*&a^*&b\\
c&b&b&a\\
\end{array}
\right),
\end{eqnarray}
with
\begin{eqnarray*}
\left\{
\begin{array}{l}
a=\cos(t\Omega/\hbar)-iE_{12}\sin(t\Omega/\hbar)/\Omega+\exp(-itE_{12}/\hbar),\nonumber\\
b=iE_J\sin(t\Omega/\hbar)/\Omega,\,\,\Omega=(E_J^2+E_{12}^2)^{1/2},\nonumber\\
c=\cos(t\Omega/\hbar)-iE_{12}\sin(t\Omega/\hbar)/\Omega-\exp(-itE_{12}/\hbar).
\end{array}
\right.
\end{eqnarray*}
The subindex ``co" refers to ``co-resonance". Thus, we can
simultaneously flip the two qubits,\, i.e.,
$|00\rangle\rightleftarrows|11\rangle$, and
$|01\rangle\rightleftarrows|10\rangle$, by setting the duration as
$\cos(t\Omega/\hbar)=-\cos(tE_{12}/\hbar)=1$. Another specific
two-qubit quantum operation
\begin{eqnarray}
\hat{U}_{\rm co}=\frac{1}{2}\left(
\begin{array}{cccc}
1-i&0&0&1+i\\
0&1+i&1-i&0\\
0&1-i&1+i&0\\
1+i&0&0&1-i\\
\end{array}
\right)
\end{eqnarray}
can also be implemented, if the duration is set as
$\cos(t\Omega/\hbar)=\sin(tE_{12}/\hbar)=1$.

\subsection{Conditional rotations of a selected qubit}

Without the interaction free subspaces~\cite{Zhou02}, a strict
single-qubit gate cannot, in principle, be achieved in the system
with strong fixed interbit-coupling. Recently, we have proposed an
effective approach to approximately implement expected
single-qubit logic operations~\cite{wei04-arXiv}. In what follows
we show that {\it conditional} single-qubit operations, i.e.,
evolving only one selected qubit and leaving the other one
unaffected, are still possible. For example, one can set
$E_C^{(k)}=E_J^{(k)}=0$ to only rotate the $j$th qubit. Indeed,
the reduced Hamiltonian $
\hat{H}_{CJ}^{(j)}=E_C^{(j)}\sigma_z^{(j)}/2
-E_J^{(j)}\sigma_x^{(j)}/2+E_{12}\,\sigma_z^{(1)}\sigma_z^{(2)}$
yields the following time-evolution
\begin{eqnarray}
\bar{U}_{\rm CJ}^{(j)}=\hat{A}_{+}^{(j)}\otimes |0_k\rangle\langle
0_k|+\hat{A}_{-}^{(j)}\otimes |1_k\rangle\langle 1_k|,
\end{eqnarray}
with
\begin{eqnarray*}
\left\{
\begin{array}{l}
\hat{A}_{\pm}^{(j)}=\mu_{\pm}^{(j)}|0_j\rangle\langle
0_j|+\mu_{\pm}^{(j)*}|1_j\rangle\langle
1_j|+\nu_{\pm}^{(j)}\sigma_x^{(j)},\\
\mu_{\pm}^{(j)}=\cos\left(t\lambda_{\pm}^{(j)}/\hbar\right)
-i\cos\alpha_{\pm}^{(j)}\sin\left(t\lambda_{\pm}^{(j)}/\hbar\right),\\
\nu_{\pm}^{(j)}=i\sin\alpha_{\pm}^{(j)}\sin\left(t\lambda_{\pm}^{(j)}/\hbar\right),\,\,
\sin\alpha_{\pm}^{(j)}=E_J^{(j)}/(2\lambda_{\pm}^{(j)}),\\
\lambda_{\pm}^{(j)}=\sqrt{[E_C^{(j)}/2\pm
E_{12}]^2+[E_J^{(j)}/2]^2}.
\end{array}
\right.
\end{eqnarray*}
This implies that, if the $k$th qubit is in the state
$|0_k\rangle$\,\, ($|1_k\rangle$), then the $j$th qubit undergoes
a rotation $\hat{A}_+^{(j)}$\,\, ($\hat{A}_-^{(j)}$). During this
operation the $k$th qubit is unchanged and kept in its initial
state.
Obviously, if  $E_C^{(j)}=2E_{12}$ is satisfied beforehand (thus
$\cos\alpha_-^{(j)}=0$), and the duration is set as
$\cos(t\lambda_j/\hbar)=1,\,\lambda_j=[(2E_{12})^2+(E_J^{(j)}/2)]^{1/2}$,
then the following two-qubit Deutsch gate~\cite{Barenco}
\begin{eqnarray}
\hat{U}_{+}^{(j)}(\theta_j)&=&\hat{I}_j\otimes |0_k\rangle\langle
0_k|\nonumber\\
&+&[\hat{I}_j\cos\theta_j+i\sigma_x^{(j)}\sin\theta_j]|1_k\rangle\langle
1_k|,
\end{eqnarray}
with $\theta_j=tE_J^{(j)}/(2\hbar)$, is obtained. Above,
$\hat{I}_j$ is the unit operator relating to the $j$th qubit. The
above operation implies that the target qubit (here it is the
$j$th one) undergoes a quantum evolution, only if the control
qubit (here, the $k$th one) is in the logical state ``1". If the
duration is set to simultaneously satisfy the two conditions:
$\sin\theta_j=1$ and $\cos(t\lambda_j/\hbar)=1$, then the above
two-qubit operation is equivalent to the well-known controlled-NOT
(CNOT) gate, apart from a phase factor. On the other hand, if
$E_C^{(j)}=-2E_{12}$ is set beforehand, then the target qubit
undergoes the same evolution only if the control qubit is in the
logic state ``0". The corresponding time-evolution operator reads
\begin{eqnarray}
\hat{U}_{-}^{(j)}(\theta_j)&=&\hat{I}_j\otimes |1_k\rangle\langle
1_k|\nonumber\\
&+&[\hat{I}_j\cos\theta_j+i\sigma_x^{(j)}\sin\theta_j]|0_k\rangle\langle
0_k|.
\end{eqnarray}

Furthermore, if $E_{C}^{(1)}=E_{C}^{(2)}=E_J^{(k)}=0$ is set
beforehand, then the above conditional operation (7) on the $j$th
qubit (keeping the $k$th one unchanged) reduces to
\begin{eqnarray}
\bar{U}_{\rm J}^{(j)}=\hat{B}_j\otimes |0_k\rangle\langle
0_k|+\hat{B}_j^*\otimes |1_k\rangle\langle
1_k|+\xi_j\sigma_x^{(j)}\otimes\hat{I}_k,
\end{eqnarray}
with
\begin{eqnarray*}
\left\{
\begin{array}{l}
\hat{B}_j=\zeta_j|0_j\rangle\langle
0_j|+\zeta_j^*|1_j\rangle\langle
1_j|,\\
\zeta_j=\cos(t\gamma_j/\hbar)
-i\cos\alpha_j\sin(t\gamma_j/\hbar),\\
\xi_j=i\sin\alpha_j\sin(t\gamma_j/\hbar),\,\,
\cos\alpha_j=E_{12}/\gamma_j,\\
\gamma_j=\sqrt{(E_{12})^2+(E_J^{(j)}/2)^2}.
\end{array}
\right.
\end{eqnarray*}
This operation can be further engineered to
\begin{eqnarray}
\hat{U}_{\rm
J}^{(j)}\hspace{-1mm}=\hspace{-1mm}\frac{i}{\sqrt{2}}\left[-\sigma_z^{(j)}\sigma_z^{(k)}
+\sigma_x^{(j)}\hspace{-0.5mm}\otimes\hspace{-0.5mm}\hat{I}_k\right],
\end{eqnarray}
if $E_J^{(j)}=2E_{12}$ and $\sin(\gamma_jt/\hbar)=1$ are further
set. This is a Hadamard-like operation on the $j$th qubit.

Of course, the above operations, although only evolve the selected
qubit and leave the other one unaffected, are no the strict
single-qubit quantum gates (but just the especial two-qubit
quantum operations). This is because the rotations of the selected
qubit depend on the states of the other one.
Note that, due to the presence of the constant interbit-coupling
$E_{12}$, the value of $E_C^{(j)}$ depends on both gate-voltages
applied to the two Cooper-pair boxes. For example, $E_C^{(2)}=0$
requires that the two gate-voltages should be set to satisfy the
condition: $(n_{g_2}-1/2)/(n_{g_1}-1/2)=-2E_{12}/E_{C_2}$.

\section{EPR-Bell states: their generations and
measurements}

Now, it will be shown how to deterministically generate EPR
correlations between the above two capacitively coupled Josephson
qubits. We will also propose how to experimentally confirm the
expected EPR-Bell states.

\subsection{Deterministic preparations}

Naturally, we begin with the ground state of the circuit
$|\psi(0)\rangle=|00\rangle$, which can be easily initialized by
letting the circuit work far from the co-resonance point via a
large voltage bias.

First, we prepare the superposition of two logical states of a
selected qubit, e.g., the first one. This can be achieved by
simply using a pulse of duration $t_1$ to implement the above
quantum operation (9), i.e.,
\begin{equation}
|\psi(0)\rangle=|00\rangle \overset{\hat{U}_-^{(1)}(\theta_1)}{\longrightarrow}%
|\Psi_{\pm}\rangle=\frac{1}{\sqrt{2}}(|00\rangle \pm i|10\rangle
).
\end{equation}
Here, the duration is set to satisfy the conditions
$\cos(t_1\lambda_1/\hbar)=1$ and $\sin\theta_1=\pm 1/\sqrt{2}$.
The plus sign corresponds to the time durations for
$\theta_1=\pi/4$, and $3\pi/4$. The minus sign corresponds to
$\theta_1=5\pi/4$, and $7\pi/4$.

We next conditionally flip the second qubit, keeping the first one
unchanged. The expected operations can be simply expressed as
either $|00\rangle\rightarrow|01\rangle$, keeping $|10\rangle$
unchanged, or $|10\rangle\rightarrow|11\rangle$, keeping
$|00\rangle$ unchanged. The former (latter) operation requires to
flip the second qubit if and only if the first qubit is in logic
state ``0" (``1"). These manipulations have been proposed above,
and thus the desirable Bell states can be deterministically
prepared by
\begin{equation}
|\Psi_{\pm}\rangle\overset{\hat{U}^{(2)}_-(\theta_2)}{\longrightarrow}|\phi
_{\pm }\rangle=\frac{1}{\sqrt{2}}(|01\rangle \pm |10\rangle),
\end{equation}
and
\begin{equation}
|\Psi_{\pm}\rangle\overset{\hat{U}^{(2)}_+(\theta_2)}{\longrightarrow}|\psi
_{\pm }\rangle=\frac{1}{\sqrt{2}}(|00\rangle \pm |11\rangle),
\end{equation}
respectively. The duration $t_2$ of the second pulse is determined
by the condition $\cos(\lambda_2t_2/\hbar)=\sin\theta_2=1$.

\subsection{Tomographic reconstructions}

The fidelity of the EPR correlations generated above can be
experimentally measured by quantum-state tomography, a technique
for reconstructing the density matrix of quantum state. For the
complete characterization of an unknown two-qubit state with a
$4\times 4$ density matrix $\rho=(\rho_{ij,kl})$ (with
$i,j,k,l=0,1$), we need to determine $15$ independent real
parameters, due to $\mbox{tr}\rho=\sum_{i,j=0,1}\rho_{ij,ij}=1$,
and $\rho_{ij,kl}=\rho_{kl,ij}^*$. This can be achieved by a
series of measurements on a sufficient number of identically
prepared copies. The operations presented above for the generation
of EPR pairs could provide enough copies of any expected EPR pairs
to be reconstructed. Experimentally, Bell states of pseudo-spins
(e.g., in nuclear magnetic resonance systems~\cite{Chuang98},
two-level trapped cold ions~\cite{Blatt04}, and the photon
pairs~\cite{White01}) have been tomographically reconstructed by
only using a series of single-qubit manipulations. Recently, we
have proposed a generic approach to tomographically measure
solid-state qubits with switchable interactions~\cite{Liu04}. Due
to the relatively strong interbit-coupling, which is always on in
the circuits considered here, specific operations are required to
realize the tomographic reconstruction of the EPR pairs generated.

The state of a charge qubit is often read out by capacitively
coupling a single-electron transistor (SET) to the measured
qubit~\cite{Tsai99}. When a projective measurement
$\hat{P}_{j}=|1_{j}\rangle \langle 1_{j}|$ is performed on the
state $\rho$, a dissipative current
$I_{c}^{(j)}\propto\mbox{tr}(\rho\hat{P}_{j})$ flows through the
$j$th SET coupled to the $j$th qubit. Such a projective
measurement is equivalent to the measurement of $\sigma
_{z}^{(j)}$, as $\sigma _{z}^{(j)}=(\hat{I}-\hat{P}_{j})/2$. For
the present system one may perform three kinds of projective
measurements: i) the $P_1$-measurement (with projective operator
$\hat{P}_{1}$) acting only on the first qubit (independent of the
state of the second qubit); ii) the $P_2$-measurement (with
projective operator $\hat{P}_{2}$) operating only on the second
qubit (independent of the state of the first qubit); and iii) the
$P_{12}$-measurement (with projective operator $\hat{P}_{1}\otimes
\hat{P}_{2}$) simultaneously acting on both Cooper-pair boxes.

All diagonal elements of the density matrix $\rho$ can be directly
determined by performing these three kinds of projective
measurements on the system. In fact, $\rho_{11,11}$ can be
determined by the $P_{12}$-measurement as
\begin{equation}
I_c^{(12)}\,\propto\,\rho_{11,11}={\rm
tr}(\rho\hat{P}_1\otimes\hat{P}_2).
\end{equation}
Next, $\rho_{10,10}$ could be determined by $P_1$-measurement as
\begin{equation}
I_c^{(1)}\,\propto\,\rho_{10,10}+\rho_{11,11}={\rm
tr}(\rho\hat{P}_1).
\end{equation}
Also, we can determine $\rho_{01,01}$ by the $P_2$-measurement as
\begin{equation}
I_c^{(2)}\,\propto\, \rho_{01,01}+\rho_{11,11}={\rm
tr}(\rho\hat{P}_2).
\end{equation}
The remaining element $\rho_{00,00}$ could be determined by the
normalization condition $\mbox{tr}\rho=1$.

The $12$ non-diagonal elements which are left, should be
transformed to the diagonal positions of new density matrix
$\rho'=\hat{W}\rho\hat{W}^\dagger$, by performing a proper quantum
operation $\hat{W}$ on the original density matrix $\rho$. For
example, after a quantum manipulation $\hat{U}_J^{(1)}$, see Eq.
(11), evolving the system to
$\overline{\rho}=\hat{U}_J^{(1)}\,\rho\,\hat{U}_J^{(1)\dagger}$,
we can perform the $P_{12}$-measurement to obtain
\begin{eqnarray}
\overline{I}_c^{(12)}&\propto&
\mbox{tr}[\overline{\rho}\,\hat{P}_1\otimes\hat{P}_2]\nonumber\\
&=&\frac{1}{2}[\rho_{01,01}+\rho_{11,11}-2\,{\rm
Re}(\rho_{01,11})],
\end{eqnarray}
for determining Re$(\rho_{01,11})$; and perform the
$P_1$-measurement to obtain
\begin{eqnarray}
\overline{I}_c^{(2)}\,\propto\,
\mbox{tr}[\overline{\rho}\,\hat{P}_1]=\frac{1}{2}[1+2\,{\rm
Re}(\rho_{00,10}-\rho_{01,11})],
\end{eqnarray}
for determining Re$(\rho_{00,10})$.
All the remaining $10$ off-diagonal elements of $\rho$ can be
similarly determined.

Table I summarizes such a procedure for tomographic
characterization of an unknown two-qubit state in this
fixed-coupling two-qubit system. We need to first apply to $\rho$
the quantum operations listed in the first column of Table I.
Afterwards, the projective measurements listed in the second
column of Table I must be made. In this way, all the matrix
elements of $\rho$ can be determined. Of course, this is not a
unique approach for determining all fifteen independent elements
of the density matrix. In fact, the expected tomographic
reconstruction could also be achieved by only using the $P_1$- and
$P_2$-measurements, and making the $P_{12}$-measurement
unnecessary.
\begin{table}
\caption{\label{tab1}Tomographic characterization of an unknown
two-qubit state $\rho=(\rho_{ij,kl})$ with $i,j,k,l=0,1$ in
capacitively-coupled Josephson circuits. Each row of this table
requires operating on an identically prepared initial state
$\rho$.}
\begin{ruledtabular}
\begin{tabular}{c|c|c}
%\cline{1-4}
 Operations & Measurement & Determining    \\
\hline
   No & $P_{12}$ & $\rho_{11,11}$ \\
\hline
   No & $P_1$ & $\rho_{10,10}$ \\
\hline
   No & $P_2$ & $\rho_{01,01}$ \\
\hline
  $\hat{U}_{\rm J}^{(1)}$& $P_{12}$ & ${\rm Re}(\rho_{01,11})$ \\
\hline
  $\hat{U}_{\rm J}^{(1)}$& $P_1$ & ${\rm Re}(\rho_{00,10})$ \\
\hline
  $\hat{U}_{\rm J}^{(2)}$& $P_{12}$ & ${\rm Re}(\rho_{10,11})$ \\
\hline
  $\hat{U}_{\rm J}^{(2)}$& $P_2$ & ${\rm Re}(\rho_{00,01})$ \\
\hline
  $\hat{U}_-^{(1)}(\frac{\pi}{4})\hat{U}_+^{(2)}(\frac{\pi}{2})$& $P_1$ & ${\rm Re}(\rho_{00,11})$ \\
 \hline
 $\hat{U}_+^{(1)}(\frac{\pi}{4})\hat{U}_+^{(2)}(\frac{\pi}{2})$& $P_{12}$ &
 ${\rm Re}(\rho_{01,10})$\\
\hline
  $\hat{U}_-^{(1)}(\frac{\pi}{4})$& $P_{2}$ & ${\rm Im}(\rho_{00,10})$ \\
\hline
  $\hat{U}_+^{(1)}(\frac{\pi}{4})$& $P_{2}$ & ${\rm Im}(\rho_{01,11})$ \\
\hline
  $\hat{U}_-^{(2)}(\frac{\pi}{4})$& $P_{2}$ & ${\rm Im}(\rho_{00,01})$ \\
\hline
 $\hat{U}_+^{(2)}(\frac{\pi}{4})$& $P_{2}$ & ${\rm Im}(\rho_{10,11})$ \\
\hline
  $\hat{U}_{\rm co}$& $P_{12}$ & ${\rm Im}(\rho_{00,11})$ \\
 \hline
  $\hat{U}_{\rm co}$& $P_{2}$ & ${\rm Im}(\rho_{01,10})$
\end{tabular}
\end{ruledtabular}
\end{table}

With the density matrix $\rho$ obtained by the above tomographic
measurements and comparing to the density matrix of ideal Bell
states, i.e.,
\begin{eqnarray*}
\rho_{|\psi_{\pm}\rangle}=\left(
\begin{array}{cccc}
1&0&0&\pm 1\\
0&0&0&0\\
0&0&0&0\\
\pm 1&0&0&1
\end{array}
\right),\, \rho_{|\phi_{\pm}\rangle}=\left(
\begin{array}{cccc}
0&0&0&0\\
0&1&\pm 1&0\\
0&\pm 1&1&0\\
0&0&0&0
\end{array}
\right),
\end{eqnarray*}
the fidelity of the EPR pairs generated above can be defined as
$F_{|\psi_{\pm}\rangle}=\mbox{tr}(\rho\rho_{|\psi_{\pm}\rangle})$
and
$F_{|\phi_{\pm}\rangle}=\mbox{tr}(\rho\rho_{|\phi_{\pm}\rangle})$,
respectively.

So far, we have shown that EPR correlations could be produced
between two capacitively coupled Cooper-pair boxes. Further, these
entangled states can be characterized by using tomographic
techniques via a series of projective measurements. Below, we will
numerically estimate the lifetimes of these states and discuss
their possible application to test Bell's inequality.

\section{Decay of EPR-Bell states due to gate-voltage noise}

The EPR pairs generated above are the eigenstates of the
Hamiltonian $\hat{H}_{\rm
int}=E_{12}\sigma_z^{(1)}\sigma_z^{(2)}$, and thus are long-lived,
at least theoretically, in the idle circuit with
$E_C^{(j)}=E_J^{(j)}=0$. Under the influence of various disturbing
perturbations, these pure quantum states will finally decay to the
corresponding mixed states. In fact, experimental solid-state
circuits are very sensitive to decoherence because of the coupling
to the many degrees of freedom of the solid-state environment.
However, coherent quantum manipulations on the generated EPR pairs
are still possible if their decay times are sufficiently long.

\subsection{Model}

The typical dominating noise in Josephson circuits is caused
either by linear fluctuations of the electromagnetic environment
(e.g., circuitry and radiation noises) or by low-frequency noise
due to fluctuations in various charge/current channels (e.g., the
background charge and critical current fluctuations). Usually, the
former one behaves as Ohmic dissipation~\cite{Weiss99} and the
latter one produces a $1/f$ spectrum~\cite{shnirman05}, which is
still not fully understood in solid-state circuits (see,
e.g.,~\cite{Gutmann03}). Here, we assume that the decay of the EPR
pairs arises from linear environmental noises, i.e., we
investigate the fluctuations of the gate voltages applied to the
qubits. Moreover, the effect of background charges that cause
dephasing are modeled by setting the zero-frequency part of the
bath spectral function to a value given by the experimentally
obtained~\cite{astafievb} dephasing rates for the charge qubit
system. This approach is valid for noise that can be approximated
as leading to an exponential decay. The effect of gate-voltage
noise on a single charge qubit has been discussed in
\cite{Weiss99}. We now study two such noises in a
capacitively-coupled circuit. Each electromagnetic environment is
treated as a quantum system with many degrees of freedom and
modeled by a bath of harmonic oscillators. Furthermore, each of
these oscillators is assumed to be weakly coupled to the
Cooper-pair boxes.

The Hamiltonian containing the fluctuations of the applied gate
voltages can be generally written as
\begin{equation*}
\widetilde{H}=\hat{H}+\hat{H}_{B}+\hat{V},
\end{equation*}
with
\begin{eqnarray}
\hat{H}_{B}=\sum_{j=1,2}\sum_{\omega _{j}}\left( \hat{a}_{\omega _{j}}^{\dagger }\hat{%
a}_{\omega _{j}}+\frac{1}{2}\right) \hbar \omega _{j},
\end{eqnarray}%
and
\begin{eqnarray}
\hat{V}=\sigma _{z}^{(1)}(X_{1}+ \beta X_{2}) + \sigma
_{z}^{(2)}(X_{2}+ \gamma X_{1}),
\end{eqnarray}%
being the Hamiltonians of the two baths and their interactions
with the two boxes. Here,
\begin{equation}
X_{j} =\frac{E_{C_{j}}C_{g_{j}}}{4e}\sum_{\omega _{j}}(g_{\omega
_{j}}^{\ast }\hat{a}_{\omega _{j}}^{\dag }+g_{\omega
_{j}}\hat{a}_{\omega _{j}}),
\end{equation}
with $\hat{a}_{\omega _{j}},\hat{a}_{\omega _{j}}^{\dagger }$
being the Boson operators of the $j$th bath, and $g_{\omega _{j}}$
the coupling strength between the oscillator of frequency $\omega
_{j}$ and the non-dissipative system. Due to the mutual coupling
of the two Cooper pair boxes, there will be crosstalk of the noise
affecting each qubit. This is modelled in the spin-boson model
with two bosonic baths represented above by the terms with the
additional factors $\beta$ and $\gamma$. The amount of this
crosstalk is given by the network of capacitances or the
corresponding energies only; namely, $\beta=E_m/2E_{C_{2}}$ and
$\gamma=E_m/2E_{C_{1}}$, and by inserting experimental values one
finds that $\beta \approx \gamma \approx 1/10$.

The effects of these noises can be characterized by their power
spectra. The spectral density of the voltage noise for Ohmic
dissipation can be expressed as
\begin{equation}
J_f(\omega)=\pi \sum_{\omega _{j}}|g_{\omega _{j}}|^{2}\delta
(\omega-\omega _{j})\, \sim \eta \hbar \omega
\omega_c^2/(\omega_c^2 +\omega^2).
\end{equation}
Here, a Drude cutoff with cutoff frequency $\omega_c=10^{4}$ GHz
has been introduced, which is well above all relevant frequency
scales of the system and given by the circuit properties
\cite{Markus03}.
%
%\begin{equation}
%\end{equation}
%
The dimensionless constant $\eta$ characterizes the strength of
the environmental effects. Introducing the impedance,
$Z_t(\omega)=1/[i\omega C_t+Z^{-1}(\omega)]$, the spectral
function for the fluctuations can be expressed via the
environmental impedance $J_f(\omega)=\omega
\mbox{Re}(Z_t(\omega))$. Here, $Z(\omega)\sim R_V$ is the Ohmic
resistor and $C_t$ is the total capacitance connected to the
Cooper-pair box.

The well-established Bloch-Redfield
formalism~\cite{AK64,Goorden03} provides a systematic way to
obtain a generalized master equation for the reduced density
matrix of the system, weakly influenced by dissipative
environments. A subtle Markov approximation is also made in this
theory such that the resulting master equation is local in time.
In the regime of weak coupling to the bath and low temperatures,
this theory is numerically equivalent to a full non-Markovian
path-integral approach~\cite{Hartmann00}. For the present case, a
set of master equations are obtained in the eigenbasis of the
unperturbed Hamiltonian~\cite{Weiss99}
\begin{equation}
\dot{\rho}_{nm}\;=\;-i\,\omega_{nm}\,\rho_{nm} -
\sum_{kl}\,R_{nmk\ell}\,\rho_{k\ell}\textrm{,}
\end{equation}
with the Redfield tensor elements $R_{nmk\ell}$ given by
\begin{equation}
R_{nmk\ell} = \delta_{\ell m}\sum_r\Gamma^{(+)}_{nrrk} +
\delta_{nk}\sum_r\Gamma^{(-)}_{\ell rrm} - \Gamma^{(-)}_{\ell mnk}
- \Gamma^{(+)}_{\ell mnk}\textrm{,}
\end{equation}
and the rates $\Gamma^{(\pm)}$ given by the Golden Rule
expressions
\begin{equation*}
\Gamma^{(+)}_{\ell mnk} = \hbar^{-2} \int_0^\infty dt\;
e^{-i\omega_{nk}t} \langle V_{I,\ell m}(t) V_{I,nk}(0)\rangle,
\end{equation*}
\begin{equation*}
\Gamma^{(-)}_{\ell mnk} = \hbar^{-2} \int_0^\infty dt\;
e^{-i\omega_{\ell m}t} \langle V_{I,\ell m}(0) V_{I,nk}(t)\rangle.
\end{equation*}
Here, $V_{I,\ell m}(t)$ is the matrix element of the system-bath
coupling term of the Hamiltonian in the interaction picture with
respect to the bath, and the brackets denote thermal average.

Note again that the strength of the dissipative effects is
characterized by the dimensionless parameter $\eta$. From
experimental measurements of the noise properties of the charge
qubit system~\cite{astafievPRL}, it is found that the strength of
the Ohmic noise is given by
\begin{equation}
\eta = \frac{4e^2 R}{\hbar \pi} \approx 1.8 \cdot 10^{-3},
\end{equation}
where $R \approx 6$ $\Omega$. Thus, current technology gives a
noise floor of approximately $\eta\sim 10^{-3}$, which will be
used for the numerical simulations.
For visualization of the decay of the Bell states, we compute the
concurrence~\cite{wootters}, given by
%$
\begin{equation}
\mathcal{C}=\mbox{max}\big \{0,\sqrt{\varrho_1}
-\sqrt{\varrho_2}-\sqrt{\varrho_3}-\sqrt{\varrho_4} \big \}.
\end{equation}
Here, the $\varrho_i,\,i=1,2,3,4$, are the eigenvalues of $\rho
\tilde \rho$ with $\tilde \rho =(\sigma_y^{1} \otimes
\sigma_y^{2}) \rho^* (\sigma_y^{1} \otimes \sigma_y^{2})$. The
concurrence is a measure for entanglement and indicates
non-locality. The maximally entangled Bell states (i.e., the ideal
EPR correlations) yield a value of $1$, whereas a fully separable
state gives $0$.

\subsection{Numerical results}

\begin{figure}[t]
\vspace{0.3cm}
% \centering
\includegraphics[width=6cm, angle=90]{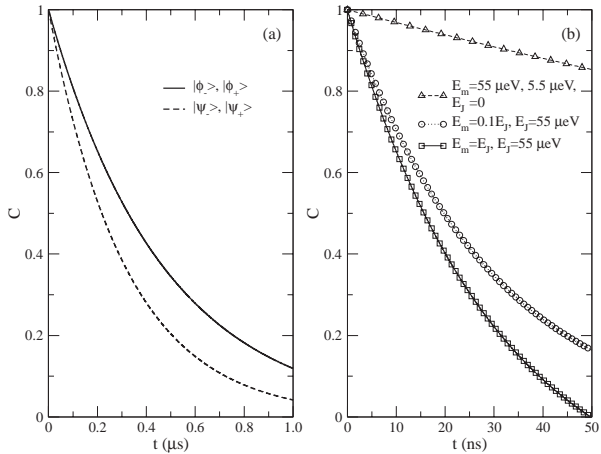}
%\vspace{-0.5cm}
\caption{Simulated time evolution of the
concurrence $\mathcal{C}$ for a two-qubit system coupled to a
noisy environment and initially prepared in the Bell states. Here,
the temperature and the strength of noise are set to $T=10$ mK and
$\eta=10^{-3}$, respectively. (a) captures the long-time decay of
the concurrence for different entangled input states in the case
of vanishing single-qubit terms, i.e., when only the inter-qubit
coupling terms are present. (b) compares the decays of
$|\psi_-\rangle$ for different interbit-couplings ($E_m=E_J$, and
$0.1E_J$) without ($E_J^{(1)}=E_J^{(2)}=0$), and with Josephson
tunneling ($E_J^{(1)}=E_J^{(2)}=E_J=55\,\mu$eV).} \label{fig:ec}
\end{figure}

The results of the simulations are shown in Fig. \ref{fig:ec},
where the time evolution of the concurrence  $\mathcal{C}$ shows
the decays of all Bell states, for temperature set to an
experimentally feasible value of $10$ mK. The lifetimes of the
operationally idle EPR pairs are of the order of several $\mu\,$s
and thus sufficiently long (compared to the duration $\sim 100$ ps
of the usual quantum manipulation).

For the case where only the coupling term between the qubits is
present and all single-qubit terms in the Hamiltonian are
suppressed, Fig. 2(a) shows that the Bell states decay
exponentially fast to zero: $\mathcal{C}(t)\,\sim\, \exp(-At)$,
with $A\simeq 2.13\times 10^{6}$ for $|\phi_{\pm}\rangle$ and
$A\simeq 3.18\times 10^{6}$ for $|\psi_{\pm}\rangle$. In this
case, only pure dephasing contributes to overall decoherence
rates, as $\hat{H}=\hat{H}_{\rm int}=E_{12}\,\sigma
_{z}^{(1)}\sigma _{z}^{(2)}$ and $[\hat{H},\hat{V}]=0$, see
Ref.~\cite{Markus03}.
The magnitude of the dephasing part of decoherence is essentially
determined by the 1/f-noise. To model this, a peak in the spectral
function at zero frequency can be introduced with a magnitude
given by microscopic calculations or experimental measurements of
the magnitude of $1/f$-noise in these qubit structures. However,
note that often the noise leads to non-exponential decay, which
can neither be modeled by Bloch-Redfield theory nor be
parametrized by a single rate. Here, we assume Markovian and
Gaussian noise and set the zero frequency contribution, i.e., the
dephasing due to the $1/f$-noise to an experimentally reported
value of $\Gamma_\varphi \approx 10^{7}$ Hz \cite{astafievb}. Note
that the individual contributions from different noise sources sum
up in the spectral function $J_\Sigma
(\omega)=J_f(\omega)+J_{1/f}(\omega)$, which also holds at
$\omega=0$. It is interesting to note that the decay time is
independent of the inter-qubit coupling strength $E_{12}$. In more
detail, when the coupling energy $E_{12}$ in the Hamiltonian is
increased the decay does not change. The reason for this behavior
is that the pure dephasing is only affected by the zero frequency
part of the spectrum, which is obviously independent of the
individual frequency splittings, i.e., the characteristic energy
scale of the Hamiltonian. Also, one of the most important results,
namely that the decay time of $|\phi_{\pm}\rangle$ is longer than
that of $|\psi_{\pm}\rangle$, is consistent with the analog
experimental one in ion traps~\cite{Blatt04}. This is because
$|\phi_{\pm}\rangle$ is the superposition of the two states with
the same energy, while $|\psi_{\pm}\rangle$ corresponds to higher
energy and is more sensitive to such perturbations.

When the Josephson-tunneling terms exist, e.g.,
$E_J^{(1)}=E_J^{(2)}=E_J$, we see from Fig. 2(b) that the decays
of the generated EPR pairs are significantly faster than in the
former case without any tunneling. This is becasue the additional
Josephson tunneling provides additional decoherence channels since
the Hamiltonian of the circuit now does not commute with the
couplings to the baths. Moreover, also the overall energy scale in
the Hamiltonian increases. In this case, the weaker
interbit-coupling corresponds to the slower decay of the EPR
pairs.

\bigskip

\section{Testing Bell's inequality}

A possible application of the deterministically generated EPR
pairs is to test BI at the macroscopic level. Due to the existence
of interbit constant-coupling, the required local operations of
encoding classical information $\{\theta_j\}$ into the EPR pairs
cannot be strictly implemented. In Ref.~\cite{wei04-arXiv} we
proposed an approach to overcome this difficulty by introducing
the effective single-qubit operations including corrections due to
the constant-coupling. Instead, here we approximately perform the
encoding procedure by sequentially applying the conditional
single-qubit operations $\bar{U}_J^{(j)},\,(j=1,2)$ in Eq. (10).
For the case of $\alpha_1=\alpha_2=\alpha$, the validity of the
above quasi-local encodings could be described by the variation of
the degree of entanglement (i.e., concurrence) of the EPR pairs
\begin{equation}
\Delta
\mathcal{C}=1-\sqrt{1-[\sin(2\alpha)(1-\cos(2\varphi_1+2\varphi_2))/2]^2}\,,
\end{equation}
with $\varphi_j=2\gamma_jt/\hbar$. Obviously, $\Delta
\mathcal{C}=0$ corresponds to the ideal locality or maximal
locality. After the above encoding, we simultaneously
detect~\cite{martinis05} the populations of qubits and check if
they are in the same logic states: the excited one $|1\rangle $ or
the ground state $|0\rangle$.

Theoretically, the correlation of two local variables, $\varphi_1$
and $\varphi_2$, can be defined as the expectation value of the
operator $\hat{P}_{T}=|11\rangle \langle 11|+|00\rangle \langle
00|-|10\rangle \langle 10|-|01\rangle \langle
01|=\hat{\sigma}_{z}^{(1)}\otimes \hat{\sigma}_{z}^{(2)}$ and
reads
\begin{equation}
E(\varphi _{1},\varphi
_{2})=\cos^2\alpha+\sin^2\alpha\cos(\varphi_1+\varphi_2).
\end{equation}
Experimentally, all the above operational steps can be repeated
many times in a controllable way for various parameter sets. As a
consequence, the correlation function $E$ can be measured by
\begin{equation}
E(\varphi _{1},\varphi _{2})=\frac{N_{\mathrm{same}}(\varphi
_{1},\varphi _{2})-N_{\mathrm{diff}}(\varphi_{1},\varphi
_{2})}{N_{\mathrm{same}}(\varphi_{1},\varphi
_{2})+N_{\mathrm{diff}}(\varphi_{1},\varphi_{2})},
\end{equation}
for any pair of chosen classical variables $\varphi_{1}$ and
$\varphi _{2}$. Here, $N_{\mathrm{same}}(\varphi_{1},\varphi_{2})$
($N_{\mathrm{diff}}(\varphi_{1},\varphi_{2})$) are the number of
events with two qubits found in the same (different) logic states.
With these measured correlation functions, one can experimentally
test the BI in the present superconducting systems.

We consider the following typical set of angles:
$\{\varphi_{j},\varphi _{j}^{\prime }\}=\{-\pi /8,\,3\pi /8\}$ and
the interbit couplings $E_m=4E_{12}=E_J, E_J/10$, and $E_J/100$,
respectively. The corresponding variations $\Delta\mathcal{C}$ of
the concurrence and the correlation $E(\varphi_{1},\varphi_{2})$,
which yields the Clauser, Horne, Shimony and Holt
(CHSH)~\cite{asp82} function $f=\left\vert
E(\varphi_{1},\varphi_{2})+E(\varphi_{1}^{\prime },\varphi
_{2})+E(\varphi_{1},\varphi_{2}^{\prime })-E(\varphi_{1}^{\prime
},\varphi_{2}^{\prime })\right\vert$, are given in Table II.
\begin{table}
\caption{\label{tab1} Variations of the concurrence,
$\Delta\mathcal{C}$, correlations $E$, and CHSH-functions $f$, for
certain typical parameters of the interbit coupling $E_m$ and the
controllable classical variables $\varphi_1$ and $\varphi_2$.}
\begin{ruledtabular}
\begin{tabular}{c|c|c|c|c}
%\cline{1-4}
 $E_m$ & $(\varphi_1,\varphi_2)$ & $\Delta\mathcal{C}$ & $E(\varphi_{1},\varphi_{2})$ & $f$\\
\hline
  & $(-\pi/8,-\pi/8)$& $0.00699$ & $0.76569$& \\
\cline{2-4}
      $E_J$ &$(-\pi/8,3\pi/8)$& $0.00699$ & $0.76569$&$2.6627$ \\
\cline{2-4}
       &$(3\pi/8,-\pi/8)$& $0.00699$ & $0.76569$& \\
\cline{2-4}
        &$(3\pi/8,3\pi/8)$& $0.26943$ & $-0.36569$& \\
\hline
 & $(-\pi/8,-\pi/8)$& $0.00238$ & $0.72434$& \\
\cline{2-4}
      $E_J/10$ &$(-\pi/8,3\pi/8)$& $0.00011$ & $0.70784$&$2.8264$
 \\
\cline{2-4}
       &$(3\pi/8,-\pi/8)$& $0.00011$ & $0.70784$& \\
\cline{2-4}
        &$(3\pi/8,3\pi/8)$& $0.00363$ & $-0.70285$&\\
\hline & $(-\pi/8,-\pi/8)$& $0.00001$ & $0.70711$& \\
\cline{2-4}
      $E_J/100$ &$(-\pi/8,3\pi/8)$& $0.00001$ & $0.70711$&$2.8284$
 \\
\cline{2-4}
       &$(3\pi/8,-\pi/8)$& $0.00001$ & $0.70711$& \\
\cline{2-4}
        &$(3\pi/8,3\pi/8)$& $0.00004$ & $-0.70706$&
\end{tabular}
\end{ruledtabular}
\end{table}
It is seen that the variations $\Delta\mathcal{C}$ of the
concurrence, after the above quasi-local operations
$\bar{U}_J^{(j)}$, decrease with decreasing interbit coupling. For
very weak coupling, e.g., $E_m/E_J=0.1\,({\rm or}\,0.01)$, the
applied conditional single-qubit operations can be regarded as
local, away from $~0.4\%$, (or $0.004\%$). Besides these tiny
loopholes of locality, Table II shows that the CHSH-type Bell's
inequality~\cite{asp82}
\begin{equation}
f(|\psi _{+}^{\prime }\rangle )<2
\end{equation}
is obviously violated.

\section{Discussion and Conclusion}

Similar to other theoretical schemes (see, e.g.,
Ref.~\cite{makhlin99}) the realizability of the present proposal
also faces certain technological challenges, such as the rapid
switching of the charge- and Josephson energies of the SQUID-based
qubits and decoherence due to the various environmental noises.
Our numerical results, considering various typical fluctuations,
showed that the lifetime of the generated EPR pairs adequately
allows to perform the required operations for experimentally
testing Bell's inequality. Indeed, for current
experiments~\cite{Pashkin03}, the decay time of a {\it two}-qubit
excited state is as long as $\sim 0.6$ ns, even for the very
strong interbit coupling, e.g., $E_m\simeq E_J$. Longer
decoherence times are possible for weaker interbit couplings. In
addition, for testing this, the influence of the environmental
noises and operational imperfections is not fatal, as the nonlocal
correlation $E(\varphi_i,\varphi_j)$ in Bell's inequality is
statistical --- its fluctuation could be effectively suppressed by
the averages of many repeatable experiments.

In summary, for the experimentally realized capacitively coupled
Josephson nanocircuits, we found that several typical two-qubit
quantum operations (including simultaneously flipping the two
qubits and only evolving a selected qubit in the case of leaving
the other one unchanged) could be easily implemented by properly
setting the controllable parameters of circuits, e.g., the applied
gate voltages and external fluxes. As a consequence of this,
macroscopic EPR correlated pairs could be deterministically
generated from the ground state $|00\rangle$ by two conditional
single-qubit operations: prepare the superposition of the two
logic states of a selected qubit, and then only flip one of the
two qubits. To experimentally confirm the proposed generation
schemes, we also propose an effective tomographic technique for
determining all density matrix elements of the prepared states by
a series of quantum projective measurements. The deterministically
generated EPR pairs provide an effective platform to test, at the
macroscopic level, certain fundamental principles, e.g., the
non-locality of quantum entanglement via violating the Bell's
inequality.

The approach proposed here can be easily modified to engineer
quantum entanglement in other ``fixed-interaction" solid-state
systems, e.g., capacitively (inductively) coupled Josephson phase
(flux) system and Ising (Heisenberg)-spin chains.

\section*{Acknowledgments}

We acknowledge useful discussions with J.Q. You, J.S. Tsai, O.
Astafiev, S. Ashhab and F.K. Wilhelm. MJS gratefully acknowledges
financial support of RIKEN and the DFG through SFB 631. This work
was supported in part by the National Security Agency (NSA) and
Advanced Research and Development Activity (ARDA) under Air Force
Office of Research (AFOSR) contract number F49620-02-1-0334; and
also supported by the National Science Foundation grant
No.~EIA-0130383.

\end{document}